\documentclass[aps,twocolumn,prc,floatfix,showpacs,preprintnumbers,amsmath,amssymb, nofootinbib,groupedaddress]{revtex4-1}
\usepackage{amsmath}
\usepackage{amssymb}
\usepackage{wasysym}
\usepackage{color}
\usepackage{graphicx}
\usepackage{dcolumn}    
\usepackage{multirow}
\usepackage{bm}         
\usepackage{sidecap}
\usepackage{ulem}
\usepackage{feynmp}
\def \beq {\begin{equation}}
\def \eeq {\end{equation}}
\def \beqa {\begin{eqnarray}}
\def \eeqa {\end{eqnarray}}
\def\d{{\rm d}}

\def\me{m_{\rm e}}

\begin{document}
%
%
\title{Theoretical calculations for precision polarimetry based on Mott scattering}

\author{X. Roca-Maza}
\email{xavier.roca.maza@mi.infn.it}
\affiliation{Dipartimento di Fisica, Universit\`a degli Studi di Milano and INFN,  Sezione di Milano, 20133 Milano, Italy}

\date{\today}

\begin{abstract}
Electron polarimeters based on Mott scattering are extensively used in different fields in physics such as atomic, nuclear or particle physics. This is because spin-dependent measurements gives additional information on the physical processes under study. The main quantity that needs to be understood in very much detail, both experimentally and theoretically, is the spin-polarization function, so called analyzing power or Sherman function. A detailed theoretical analysis on all the contributions to the effective interaction potential that are relevant at the typical electron beam energies and angles commonly used in the calibration of the experimental apparatus is presented. The main contribution leading the theoretical error on the Sherman function is found to correspond to radiative corrections that have been qualitatively estimated to be below the 0.5\% for the considered kinematical conditions: unpolarized electron beams of few MeV elastically scattered from a gold and silver targets at backward angles.     
\end{abstract}



\maketitle 

\section{Introduction}
\label{intro}

Electron polarimeters based on Mott scattering are extensively used in different fields in physics such as atomic, nuclear or particle physics \cite{gay1992}. This is because spin-dependent measurements gives additional information on the physical processes under study \cite{kessler1985}. Mott polarimeters are instrumental in the study of spin-dependent effects in atomic collisions, magnetization in solids, parity violating in low-energy nuclear and high-energy particle physics, precision measuremets of the $Z^0$ mass or tests of special relativity \cite{gay1992}. Nowadays, due to the advent of new  experimental programs at JLab \cite{jlab} and MAMI/MESA \cite{mesa} facilities where accuracy on electron spin-polarization will be crucial in different experiments, it is timely to provide state-of-the-art theoretical calculations of the Sherman function at the typical kinematical conditions and target nucleus of interest for the calibration of the Mott polarimeters.    

Elastic collisions of electron projectiles with atoms and ions are usually described by means of the static-field approximation \cite{mott65}. Exchange effects can be approximately accounted for by adding an approximate local-exchange interaction \cite{bransden76} to the electrostatic potential. The accuracy of this approximation is limited by inelastic absorption and charge-polarization effects. The existence of open inelastic channels implies a loss of projectile flux from the elastic channel\footnote{For electron energies of the order of few MeV the absorptive and correlation-polarization effects become negligible.}. Within these approximations, the interaction potential is completely determined by the adopted nuclear and electronic charge-densities. The nuclear charge density of the target nucleus can be modeled by a parametrized function fitted to experiment or derived from microscopic models such as those based on effective interactions solved within the mean-field approach \cite{roca08,roca13}. The latter framework has been shown to be reasonable in the description of bulk nuclear properties along the whole nuclear chart \cite{bender03}. For incident electron energies of some MeV, the two approaches yield equivalent scattering observables. The most accurate electronic charge-densities available for free neutral atoms are obtained from the multiconfiguration Dirac-Fock code of Desclaux \cite{desclaux75}.

A theoretical description of high-precision measurements on electron scattering observables, may require to also account for the so called radiative corrections \cite{johnson1962,motz1964}. Radiative corrections are calculated from processes involving real photons ({\it soft}-Bremsstrahlung) and virtual photons (QED corrections to the tree level). At lowest order, the leading QED corrections are the vacuum polarization and self-energy. For heavy nuclei with a large charge number $Z$ these QED effects cannot be treated perturbatively \cite{soff88, beier99, shabaev00}. The self-energy and vacuum-polarization contributions, being of the same order in the expansion on the fine-structure constant, are expected to be of the same order of magnitude. Moreover, they are of opposite sign when evaluated on atomic electrons \cite{shabaev00}. Self-energy corrections are very complicated to be reliably evaluated with high accuracy. They introduce a non-locality in the interaction potential, require renormalization --divergent terms appear-- and, within our approximation, require to work with electron wave functions that are eigenstates of the Coulomb potential and not simple plane waves. On the other side, the vacuum-polarization correction can be more easily estimated by means of the Uehling approximation \cite{uehling35,wayne76}. 
   
The calculations presented here have been performed by using the code {\sc elsepa} \cite{salvat05} and later modifications \cite{roca08}. The last modification was the inclusion of the Uehling potential to evaluate the effect of the vacuum-polarization correction. {\sc elsepa} allows relativistic partial-wave calculations to be performed for projectiles with kinetic energies up to several MeV and for a variety of interaction potentials. Elastic differential cross sections (DCSs) and spin-polarization functions calculated in this way constitute the state-of-the-art for energies in the MeV level. 

The present article is essentially based on Refs.~\cite{salvat05,roca08,roca13,doris14} and it is organized as follows. In Sec.\ref{theory}, for the sake of completeness, a review of the theory necessary to produce the results shown in Sec.\ref{results} is briefly discussed. Our conclusions are laid in Sec.\ref{conclusions}.  


\section{Theory}
\label{theory}

The elastic interaction of an electron with a target atom of atomic number $Z$ placed at the origin of coordinates is considered. The electron cloud as well as the nuclear charge densities have been assumed to be spherically symmetric. The interaction between the electron at a distance $r$ and the target is described by means of an optical-model potential, 
\beq
V(r) = V_{\mathrm{st}}(r) + V_{\mathrm{ex}}(r) - i W_{\mathrm{abs}}(r)\ , 
\eeq
where $V_{\mathrm{st}}(r)$ is the electrostatic interaction potential, $V_{\mathrm{ex}}(r)$ is an exchange potential that accounts for the occurrence of rearrangement collisions in which the projectile electron exchanges place with an atomic electron \cite{bransden76} and $W_{\mathrm{abs}}(r)$ is the absorption potential\footnote{Correlation-polarization potential is only needed for slow projectiles ($E\leq 10$ keV) so it has been neglected here.}. Then, elastic-scattering properties can be calculated by using conventional partial-wave methods \cite{salvat05}.

The potential energy of an electron at a distance $r$ from the center of the nucleus is given by 
\beqa
V_{\mathrm{st}}(r) &\equiv &-e\varphi(r)\equiv -e[\varphi_{\mathrm{nucl.}}(r)+\varphi_{\mathrm{at. elec.}}(r)] \\
&=&- 4\pi e \left(\frac{1}{r} \int_0^r \rho_{\rm n} (r') \, r'^2 \d r' \right.
+ \left. \int_r^\infty \rho_{\rm n}(r') \, r' \d r'\right) \nonumber\\
&&-4\pi e \left(\frac{1}{r} \int_0^r \rho_{\rm e} (r') \, r'^2 \d r' \right.
+ \left. \int_r^\infty \rho_{\rm e}(r') \, r' \d r'\right)\nonumber
\label{cesc.1}
\eeqa
where $\rho_{\rm n}(r)$ denotes the charge density of the nucleus normalized to $Z$ and $\rho_e(r)$ that of atomic electrons normalized to $Z$ for neutral atoms---and to the total number of electrons for ions. At the energies of interest, the effect of screening by the orbiting atomic electrons is limited to small scattering angles. To quantify the screening of the nuclear charge by the atomic electrons, one can define the screening function, $\chi(r)\equiv\frac{r}{Ze}\varphi(r)$. Since the electrostatic potential and the particle densities of the atom follows the Poisson's equation, one finds in spherical symmetry a relation between the nuclear and electron electric charge densities and the screening function \cite{salvat05}.

The DCS for elastic scattering of spin unpolarized electrons is given by 
\beq
\frac{\d \sigma}{\d \Omega} = |f(\theta)|^2 + |g(\theta)|^2 ,
\label{els.8}\eeq
where
\begin{eqnarray}
f(\theta) = \frac{1}{2 {\rm i} k} \sum_{\ell=0}^\infty &\bigg\{& 
(\ell+1) \left[ \exp\left( 2 {\rm i} \delta_{\kappa=-\ell-1} \right) -
1 \right] \nonumber \\
&+& \ell \left[ \exp \left( 2 {\rm i} \delta_{\kappa=\ell} \right) - 1
\right]
\bigg\} \, P_\ell(\cos\theta) 
\label{els.2}
\end{eqnarray}
and 
\beqa
g(\theta) = \frac{1}{2 {\rm i} k}
\sum_{\ell=1}^\infty &\bigg[& \exp \left( 2 {\rm i}
\delta_{\kappa=\ell}\right) \nonumber \\
&-& \exp\left( 2 {\rm i} \delta_{\kappa=-\ell-1} \right)
\bigg] \, P_\ell^1(\cos\theta)
\label{els.3}\eeqa
are the direct and spin-flip scattering amplitudes, respectively; $k$ denotes the wave number of the projectile electron $c \hbar k =  \sqrt{E(E+2\me c^2)}$; and the functions $P_\ell(\cos\theta)$ and $P_\ell^1(\cos\theta)$ are Legendre polynomials and associated Legendre functions, respectively. The phase shifts $\delta_\kappa$ represent the behavior of the Dirac spherical waves at large $r$ distances. Relative numerical uncertainties of the computed scattering amplitudes and DCS are estimated from the convergence of the partial-wave series, they are typically smaller than $10^{-6}$. {\sc elsepa} \cite{salvat05} also provides the Sherman function---or analyzing power,
\beq
S(\theta)\equiv \imath \frac{f(\theta )g^*(\theta)-f^*(\theta)g(\theta)}{\vert f(\theta)\vert^2 + \vert g(\theta)\vert^2}
\eeq
which gives the degree of spin polarization of the electrons from an initially unpolarized beam that are scattered in the direction $\theta$.

Also important are the spin rotation functions $T(\theta)$ and $U(\theta)$, 
\beqa
T(\theta)&\equiv& \frac{\vert f(\theta)\vert^2 - \vert g(\theta)\vert^2}{\vert f(\theta)\vert^2 + \vert g(\theta)\vert^2}\\
U(\theta)&\equiv& \frac{f(\theta )g^*(\theta)+f^*(\theta)g(\theta)}{\vert f(\theta)\vert^2 + \vert g(\theta)\vert^2} \ .
\eeqa
The square of the spin functions should fulfill the sum rule $S(\theta)^2 + T(\theta)^2 + U(\theta)^2 = 1$. 

The spherical solutions of the Dirac equation are suitably expressed in the form \cite{salvat05}
\beq
\psi_{E\kappa m}({\bf r}) = \frac{1}{r}
\left( \begin{array}{c}
P_{E\kappa}(r) \, \Omega_{\kappa, m} (\hat{\bf r}) \\ [1mm]
{\rm i} Q_{E\kappa}(r) \, \Omega_{-\kappa, m} (\hat{\bf r})
\end{array} \right) \ ,
\label{els.5}\eeq
where the functions $\Omega_{\kappa, m} (\hat{\bf r})$ are the spherical spinors, and the radial functions $P_{E\kappa}(r)$ and $Q_{E\kappa}(r)$ satisfy the system of coupled differential equations
\beqa
\frac{\d P_{E\kappa}}{\d r} &=&
-\frac{\kappa}{r} P_{E\kappa}
+\frac{E-V+2\me c^2}{c\hbar} \, Q_{E\kappa},
\nonumber \\ [4mm]
\frac{\d Q_{E\kappa}}{\d r} &=&
- \frac{E-V}{c\hbar} \, P_{E\kappa}
+ \frac{\kappa}{r} Q_{E\kappa}.
\label{els.6}\eeqa
$\kappa$ is defined as $\kappa = (\ell-j)(2j+1)$, where $j$ and $\ell$ are the total and orbital angular momentum quantum numbers. For modified Coulomb potentials and $r\rightarrow\infty$: $P_{E \kappa} (r) \simeq \sin \left( kr - \ell \frac{\pi}{2} - \eta \ln 2kr + \delta_\kappa\right)$ where $\eta = Z e^2 \me /(\hbar^2 k)$ is the Sommerfeld parameter (see, e.g., Ref.\ \cite{salvat05} and references therein).

\subsection{The nuclear charge distribution}

\begin{figure}[t!]
\centering
\includegraphics[width=\linewidth,clip=true]{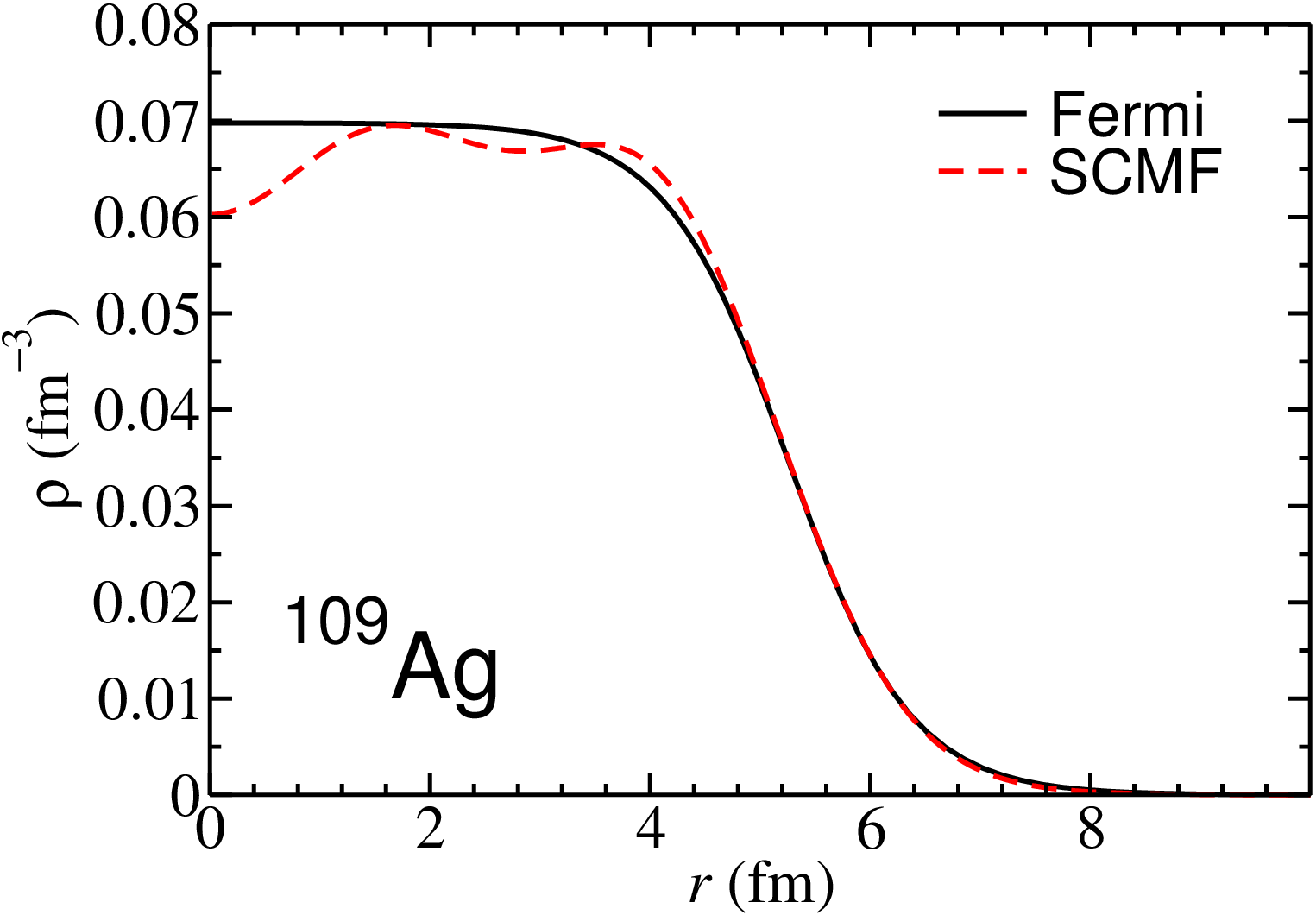}
\caption{Nuclear charge density as a function of the distance to the center for ${}^{109}$Ag. A Fermi model and a self-consistent mean-field model are shown.}
\label{den}
\end{figure}

The details of the nuclear charge distribution affect the calculated scattering observables for projectiles with kinetic energies larger than about 50 MeV. For energies of the order of the MeV, it is a good approximation to consider a simplified model such as the two parameter Fermi function or the Helm model \cite{roca08,roca13}. The nuclear charge density has been modeled by means of a two parameter Fermi function, 
\beq
\rho_{\mathrm{F}}(r) = \frac{\rho_0}{1+\exp[(r-C)/a]} 
\eeq
where $\rho_0$ is determined by fixing the charge of the nucleus ($Z$); $a$ describes the diffuseness of the surface of the density profile; and $C$ describes the mean location of this surface (i.e., $C$ is indicative of the extension of the bulk part of the density distribution). The parameters $a$ and $C$ have been determined so that they reproduce the experimental root mean square charge radius, $\langle r_{\mathrm{ch}}^2\rangle^{1/2}$ of $4.5601\pm 0.0035$ fm and $5.4371\pm 0.0038$ fm, for ${}^{109}$Ag and ${}^{197}$Au, respectively \cite{angeli04}. Specifically, the parameters of the Fermi function are $a = 0.5573$ fm and $C = 5.250$ fm for ${}^{109}$Ag and $a=0.58187$ fm and $C=6.440$ fm for ${}^{197}$Au. As an example, in Fig.\ref{den} the charge density corresponding to ${}^{109}$Ag is shown. For comparison, the prediction of a self-consistent mean-field model (SCMF \cite{lalazissis05}) is also displayed. The SCMF approach assumes that nucleons move independently in a mean field generated by the other nucleons of the atomic nucleus and it is known to be accurate for the description of bulk properties of nuclei such binding energies or charge distributions \cite{bender03}. These phenomenological models usually depend on about ten adjustable parameters that are fitted to reproduce relevant ground-state properties for some selected nuclei. For recent works that analyze the accuracy of mean field models in the description of the DCS in elastic electron scattering, the reader is referred to Ref.\cite{roca08}.

As seen in Fig.\ref{den}, the Fermi distribution displays the correct surface fall-off behavior when compared to more sophisticated calculations such as SCMF. On this regard, it has been checked that an accurate nuclear self-consistent mean-field model give the same result within 0.1 \% error in the Sherman function (cf. inset of Fig.\ref{sherman-ag}) when compared with the fitted Fermi and Helm distributions at the kinematical conditions and target nuclei of interest.  

\begin{figure}[t]
\centering
\includegraphics[width=\linewidth,clip=true]{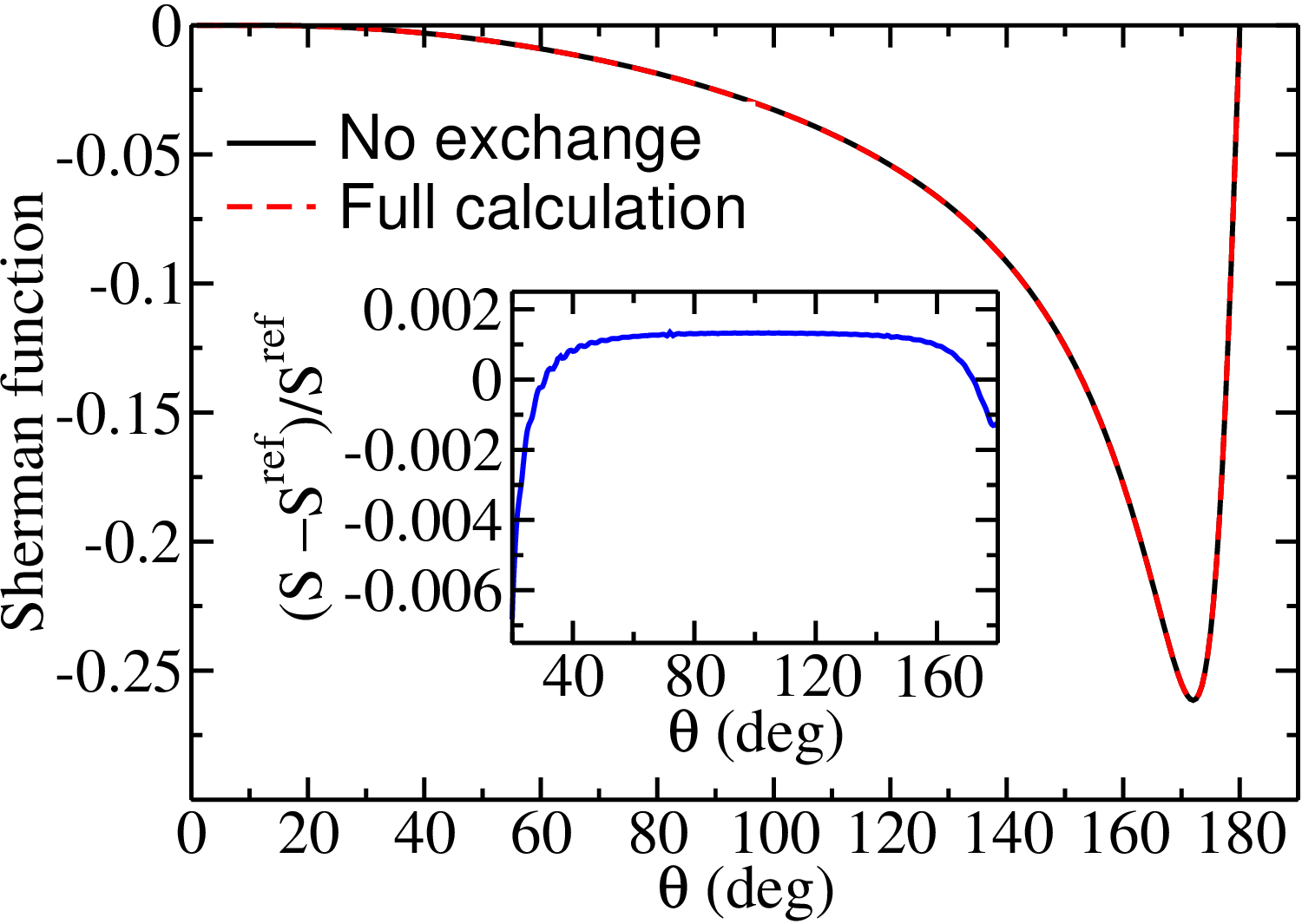}  
\caption{Sherman function of 5 MeV electrons on a ${}^{109}$Ag target as a function of the scattering angle. Calculations represented by the solid black line neglect the effect of the exchange potential while the ones represented by a dashed red line include such an effect. In the inset, the relative difference is depicted.}
\label{exch}
\end{figure}

\subsection{Electron density model}

In this work, the most accurate electron densities available for free atoms obtained from self-consistent relativistic Dirac-Fock (DF) calculations \cite{desclaux75} has been adopted. The effect in the Sherman function of neglecting atomic electrons in scattering processes of few MeV electrons may produce an error of about a few \% (cf. Fig.\ref{sherman-ag}) and, thus, the calculations presented in the next section will account for electron screening effects.

\subsection{Electron exchange potentials}

When the projectile is an electron, one should account for rearrangement collisions in which the projectile exchanges places with an atomic electron. In relativistic-electron elastic scattering under some reasonable approximations \cite{salvat05}, the scattering wave function is found to satisfy an equation similar to the Dirac-Fock equations, with a non-local exchange term. A simpler, and computationally more convenient approach is to use local approximations to the exchange interaction. Specifically, the Furness-McCarthy exchange potential \cite{furness73} has been used here. The effect in the Sherman function of neglecting the effect of the exchange potential in scattering processes of few MeV electrons may produce an error of about a few \permil  ~(cf. Fig.\ref{exch}). So negligible for our purposes.

\subsection{Inelastic contributions}

\begin{figure}[t!]
\centering
\includegraphics[width=\linewidth,clip=true]{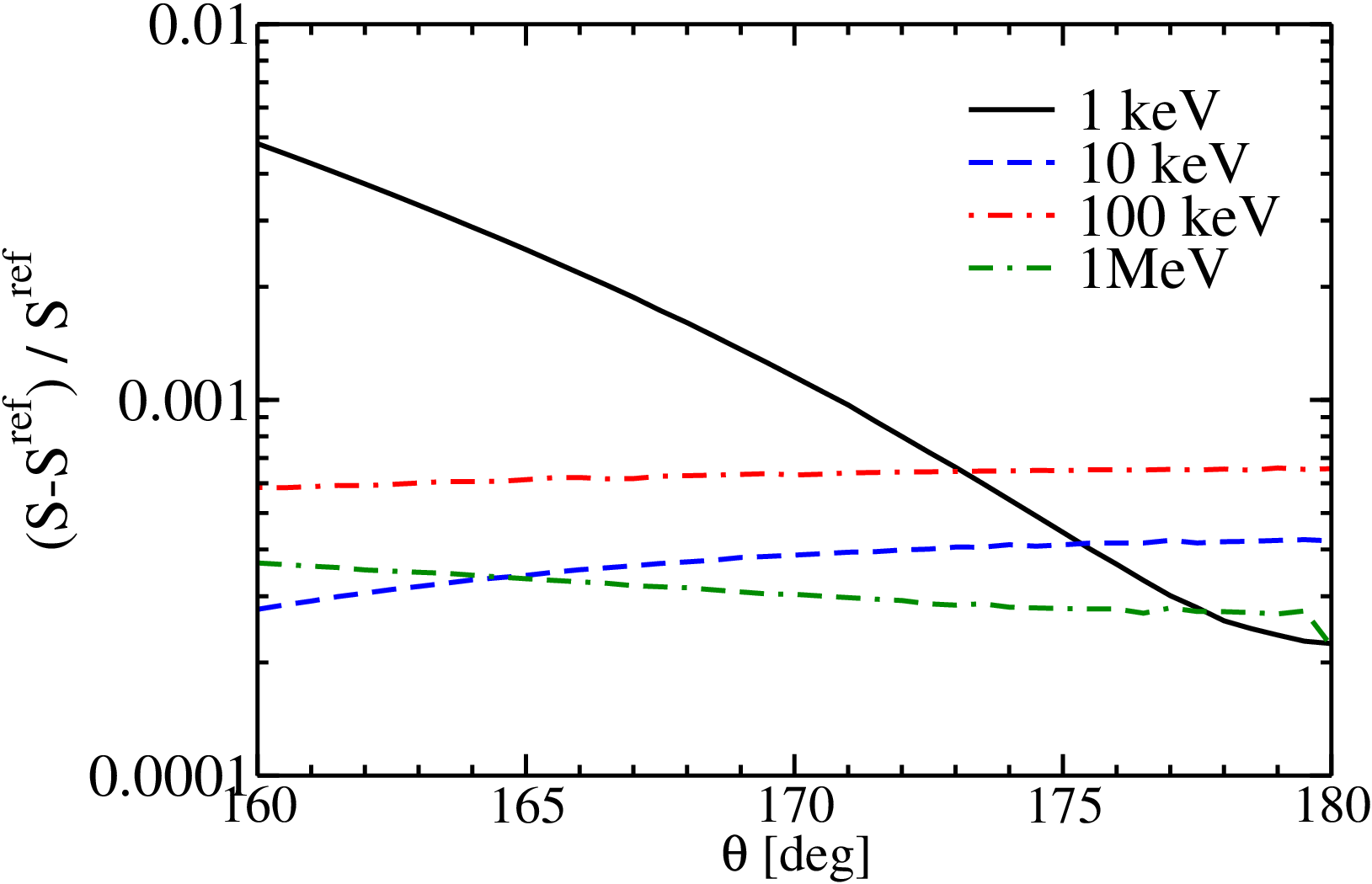}
\caption{Relative contribution from the inelastic channels to the Sherman function as a function of the angle and for different electron energies for the case of ${}^{197}$Au.}
\label{inel}
\end{figure}

The loss of particles from the elastic to the inelastic channels is modeled by including a negative imaginary term, $-iW_{\rm abs}(r)$ in the optical-model potential. The potential $W(r)$ proposed by Salvat \cite{salvat03} that is obtained by means of the Local Density Approximation is used. In very brief, it is assumed that a projectile interacts with the atomic electron cloud as if it were moving within an homogeneous electron gas. This model can be applied to energies $\leq$ 1 MeV and it can already give some hints on the effects one should expect at few MeV electron energies. In Fig.\ref{inel}, the relative effect of the inelastic channels in the Sherman function at backward angles and energies 1keV-1MeV on a ${}^{197}$Au target is shown. Hence, an error well below 0.1\% should be expected due to the inelastic channels for electron energies of few MeV at backward angles. 

\subsection{Radiative corrections}
In electron scattering by a Coulomb field, real photons are emitted. In the so-called {\it soft} (or low-energy) photon region\cite{motz1964}, inelastic processes cannot be distinguished from elastic processes. This is related with the experimental energy resolution. Radiative corrections are calculated from processes involving real photons ({\it soft}-Bremsstrahlung) and virtual photons (QED corrections to the tree level). Both types of contributions are, in principle, needed to accurately describe the experimental data. At lowest order, the leading QED corrections are the vacuum polarization and self-energy. At the kinematical conditions and target of interest, there has only been experimental evidence that radiative corrections do not amount by more than a 0.5\% in the Sherman function in the limit of zero-thickness target \cite{steigerwald}. In this experimental reference, measurements at three different electron energies between 2 and 8 MeV and gold target were performed neglecting in the analysis all radiative corrections. Since these corrections increase with energy and in Ref.\cite{steigerwald} just a single fit for all energies gave a very good accuracy, one may reasonably expect that radiative corrections do not amount by more than a 0.5\%.

\begin{figure}[t]
\centering
\includegraphics[width=\linewidth,clip=true]{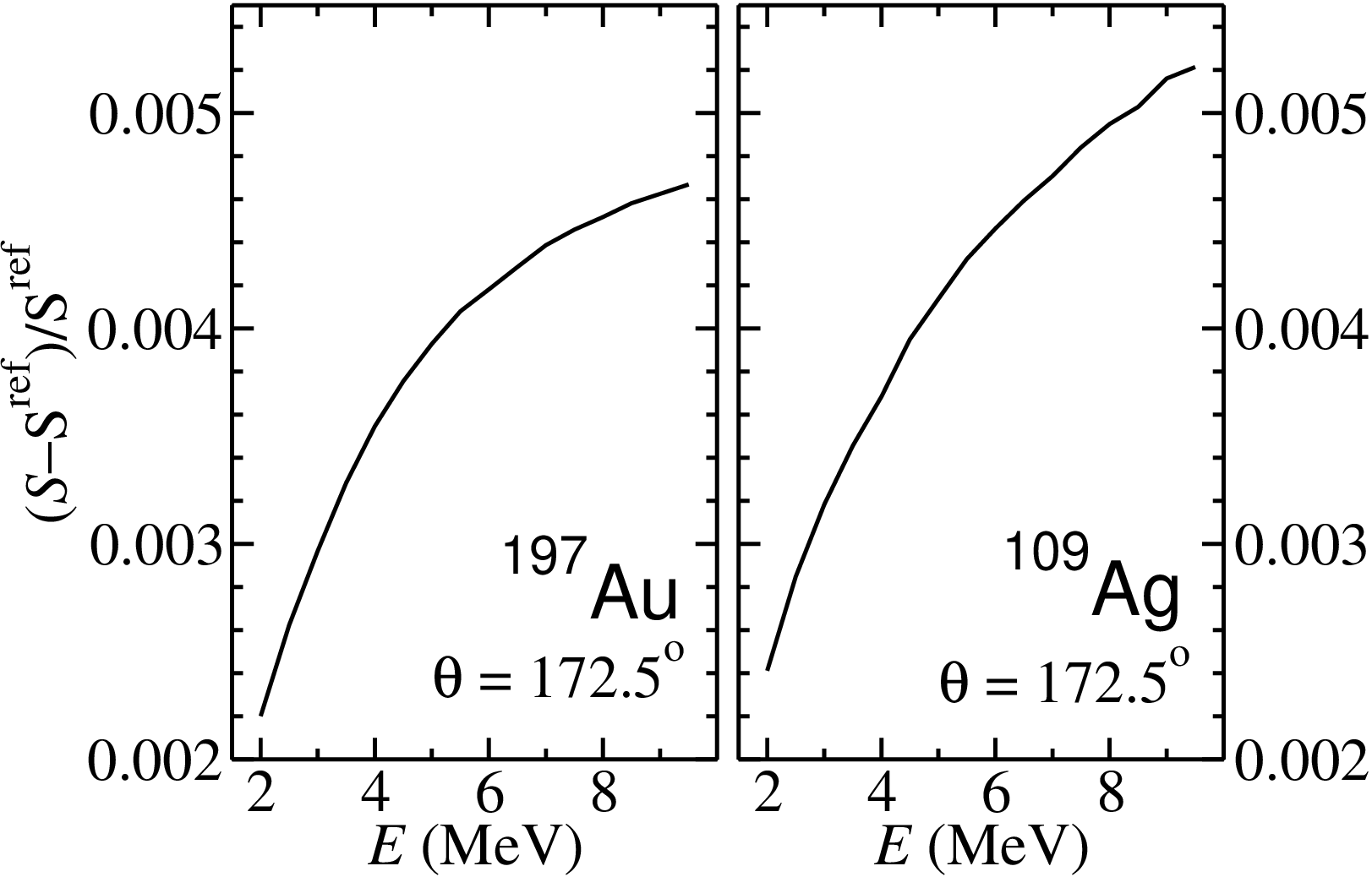}
\caption{Relative effect of vacuum polarization corrections to the Sherman function on electron beam energy at backward angles for ${}^{197}$Au (left panel) and ${}^{109}$Ag (right panel).}
\label{vac}
\end{figure}

The vacuum polarization correction is evaluated in this work by following the Uehling approximation \cite{uehling35,wayne76} but, instead of doing it perturbatively, the Dirac equation will be solved in the combined potential \cite{doris14}. That is, the potential due to the vacuum-polarization effects is added to $V(r)$. Specifically, the vacuum polarization correction is of 0.5\% or below (increases with energy) for the kinematical conditions and target of interest (cf. Fig.\ref{vac} and Ref.\cite{doris14}). 

Finally, it is important to note that it would be misleading to include such a correction in the calculations and neglect the self-energy and {\it soft}-Bremsstrahlung corrections. All radiative corrections should be included for a fully consistent calculation. So, the vacuum-polarization correction will be only used to show its impact on the theoretical predictions presented here. 

\subsection{Final remarks}

A macroscopic Bremsstrahlung effect (out of the context of radiative corrections discussed here) when many beam electrons lose energy can be simulated in real targets. Recoil effects \cite{foldy1959} might be also a relevant issue on these scattering processes. Nevertheless, these effects have been neglected in the present calculations since they can be corrected in the simulations of the experimental data. The same applies for other issues such as the real thickness of the target.

On the other side, the numerical accuracy and method discrepancies in the presented results have been checked by comparing with other available codes \cite{doris} finding that it is within the 0.1\% accuracy. Taking into account all the previous considerations, the theoretical calculations presented here might be used for the calibration of a Mott polarimeter to an accuracy of 0.5\% in the region of interest. The main theoretical error is coming from the effect of radiative corrections. 

\section{Results}
\label{results}

\begin{figure}[t]
\centering
\includegraphics[width=\linewidth,clip=true]{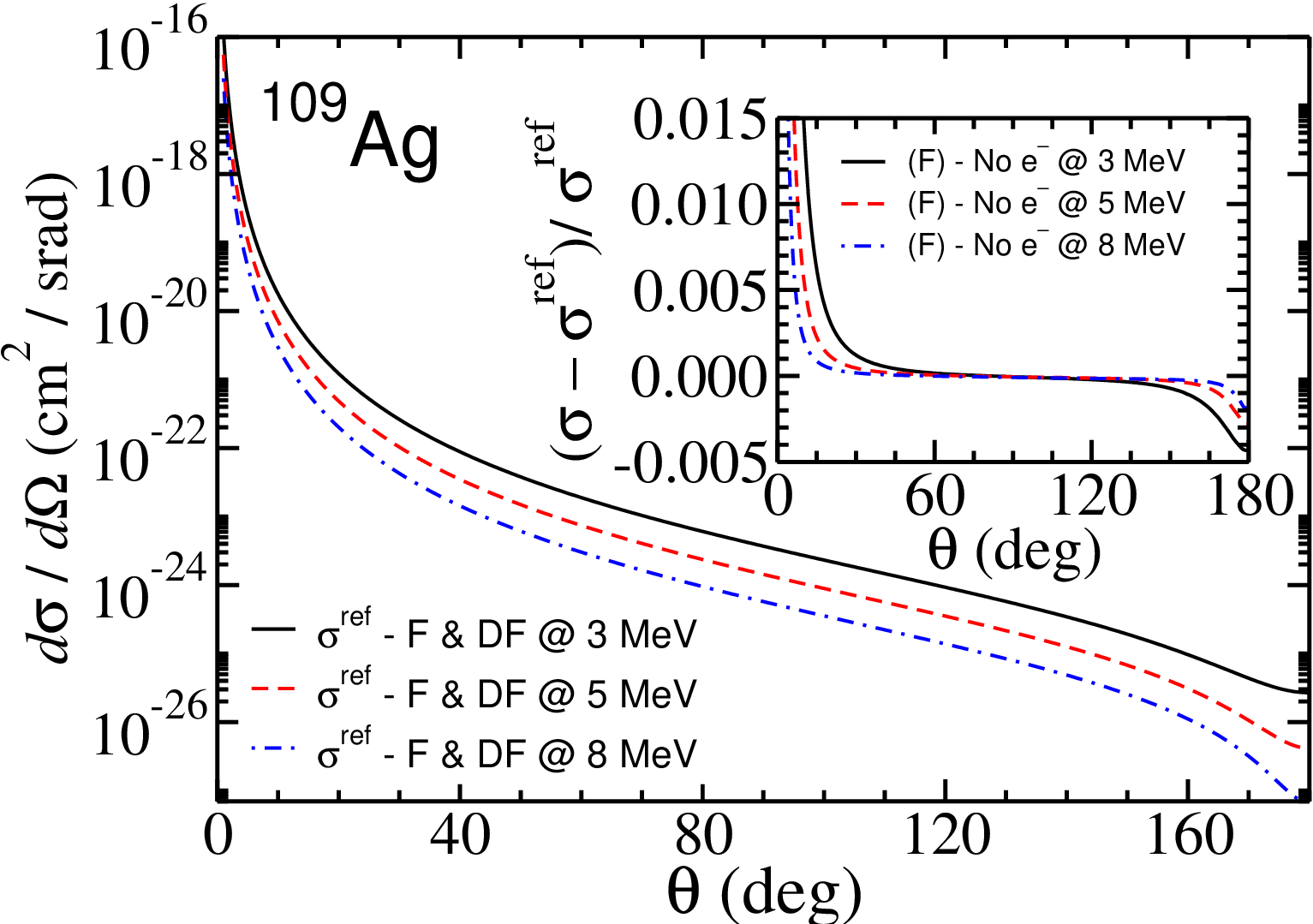}  
\includegraphics[width=\linewidth,clip=true]{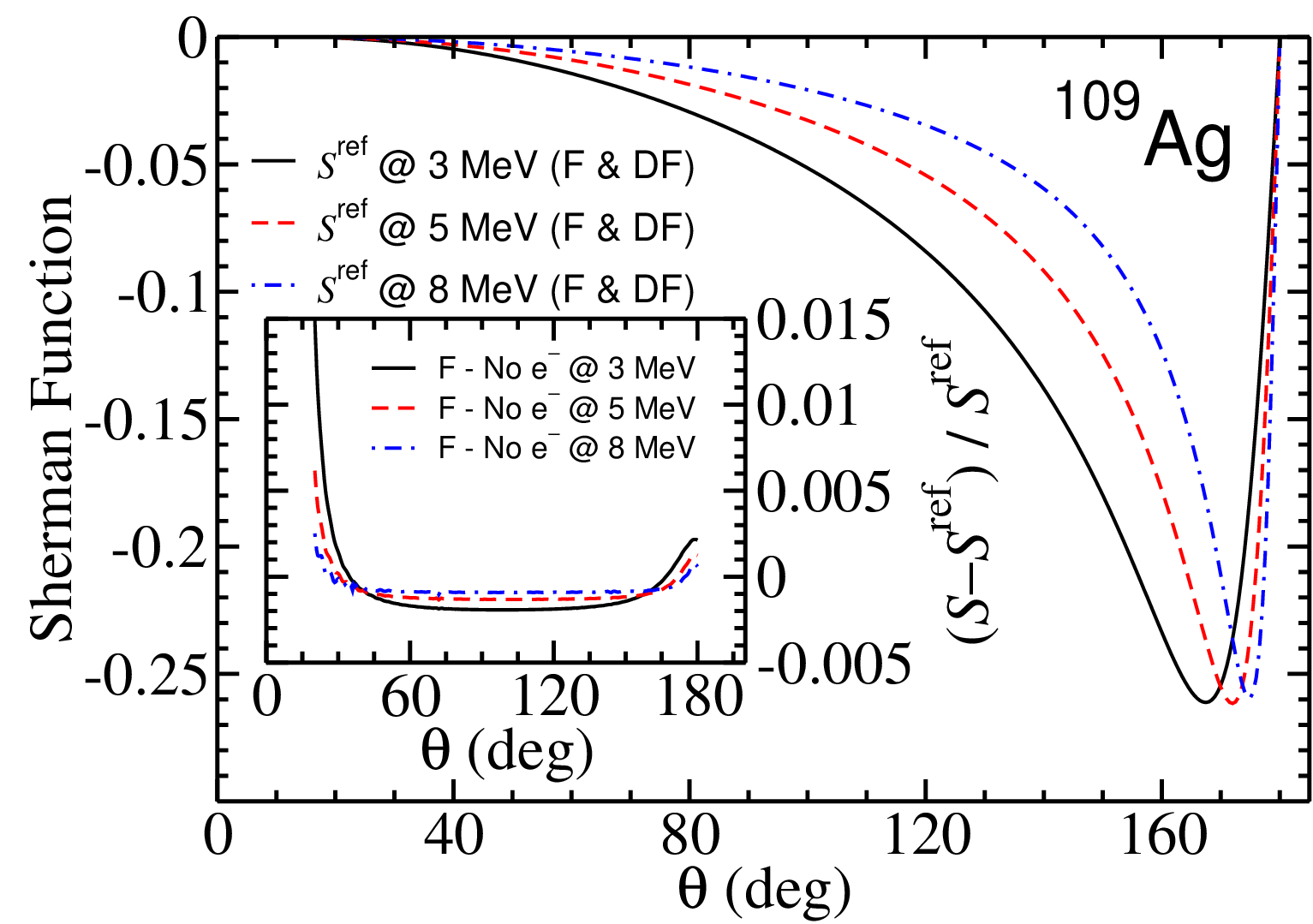}  
\caption{Elastic differential cross section (upper panel) and Sherman function (lower panel) as a function of the scattering angle for 3, 5 and 8 MeV electrons by ${}^{109}$Ag. F stands for Fermi function and the label DF indicates that the screening of the nuclear charge by the atomic electrons is included. In the insets, the relative change of the same quantity as a function of the energy with respect to the case in which the effect of atomic electrons is neglected.}
\label{dcse}
\end{figure}

In this section, some of the results obtained from calculations of the DCS and Sherman function in elastic scattering of few MeV electrons by ${}^{109}$Ag and ${}^{197}$Au are presented. First of all different test calculations described in Sec.\ref{theory} are shown. In Fig.\ref{dcse} the elastic DCS (upper panel) and the Sherman function (lower panel) as a function of the scattering angle for 3, 5 and 8 MeV electrons by  ${}^{109}$Ag are displayed. The results correspond to calculations where all ingredients have been included---except for the vacuum polarization corrections; F means that the Fermi function is used to model the nuclear charge distribution; and the label DF indicates that the screening of the nuclear charge by the atomic electrons is included. The DCS decreases with energy. In the insets, we show the relative change of the same quantities with respect to the case in which the presence of atomic electrons is neglected, which is a good approximation except for small angles.  

\begin{figure}[t]
\centering
\includegraphics[width=\linewidth,clip=true]{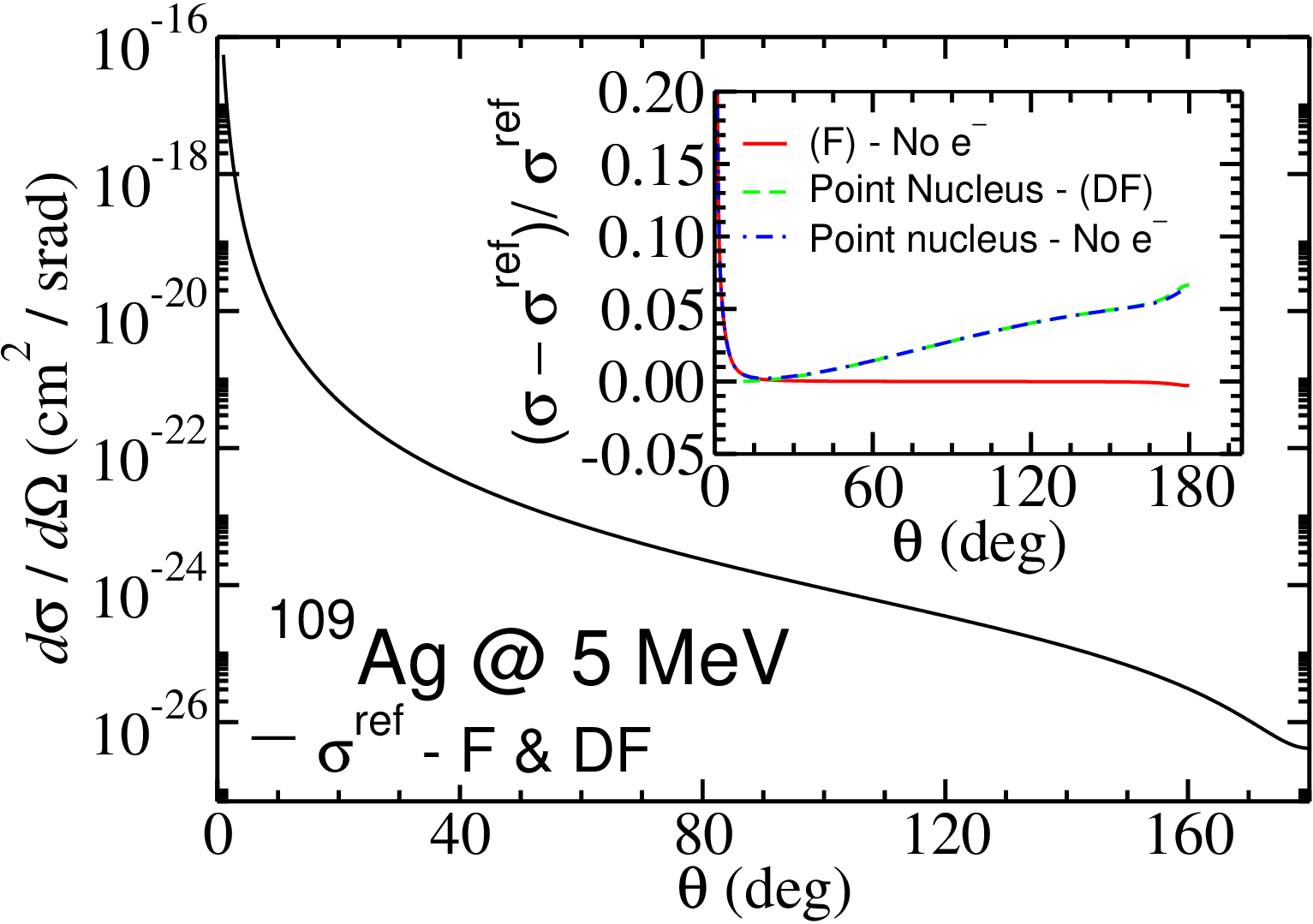}    
\includegraphics[width=\linewidth,clip=true]{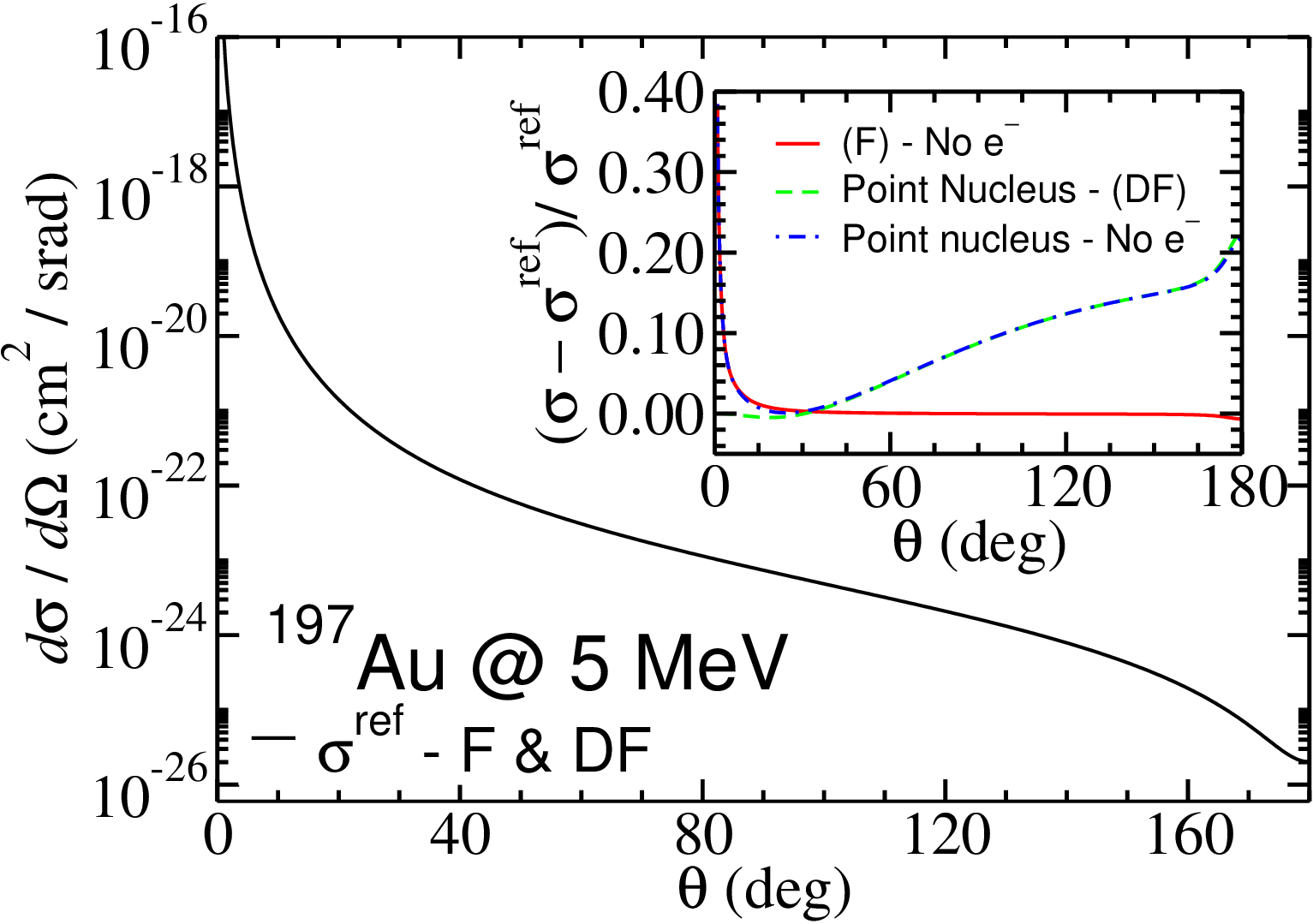}  
\caption{Elastic differential cross section as a function of the scattering angle for 5 MeV electrons by  ${}^{109}$Ag (upper panel) and ${}^{197}$Au (lower panel). In the inset, the relative change of the same quantity adopting different approximations and with respect the full calculation are displayed.}
\label{dcs}
\end{figure}

In Fig.\ref{dcs} the elastic differential cross section as a function of the scattering angle for 5 MeV electrons and ${}^{109}$Ag (upper panel) and ${}^{197}$Au (lower panel) targets are shown. The results correspond to calculations where all potential components described in Sec.\ref{theory} have been included---except for the vacuum polarization corrections. In the inset, the relative change of the same quantity adopting different approximations and with respect the full calculation is displayed: neglecting atomic electrons (red); assuming a point-like nucleus (blue); and adopting the last two approximations (green). It is clear from the inset, that neglecting the finite size of the nucleus may produce errors of a few tens of a percent in the DCS. It is also evident that the effect of atomic electrons is much less relevant.    

\begin{figure}[t]
\centering
\includegraphics[width=0.49\linewidth,clip=true]{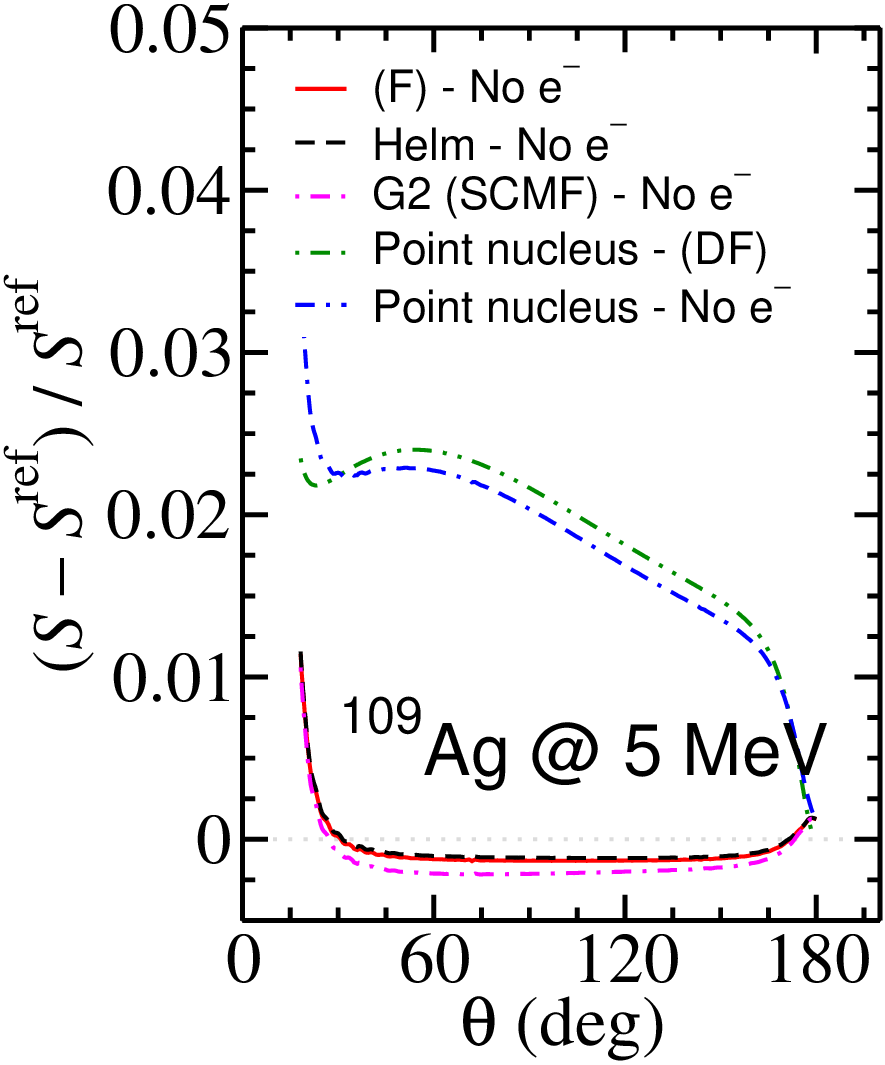}
\includegraphics[width=0.49\linewidth,clip=true]{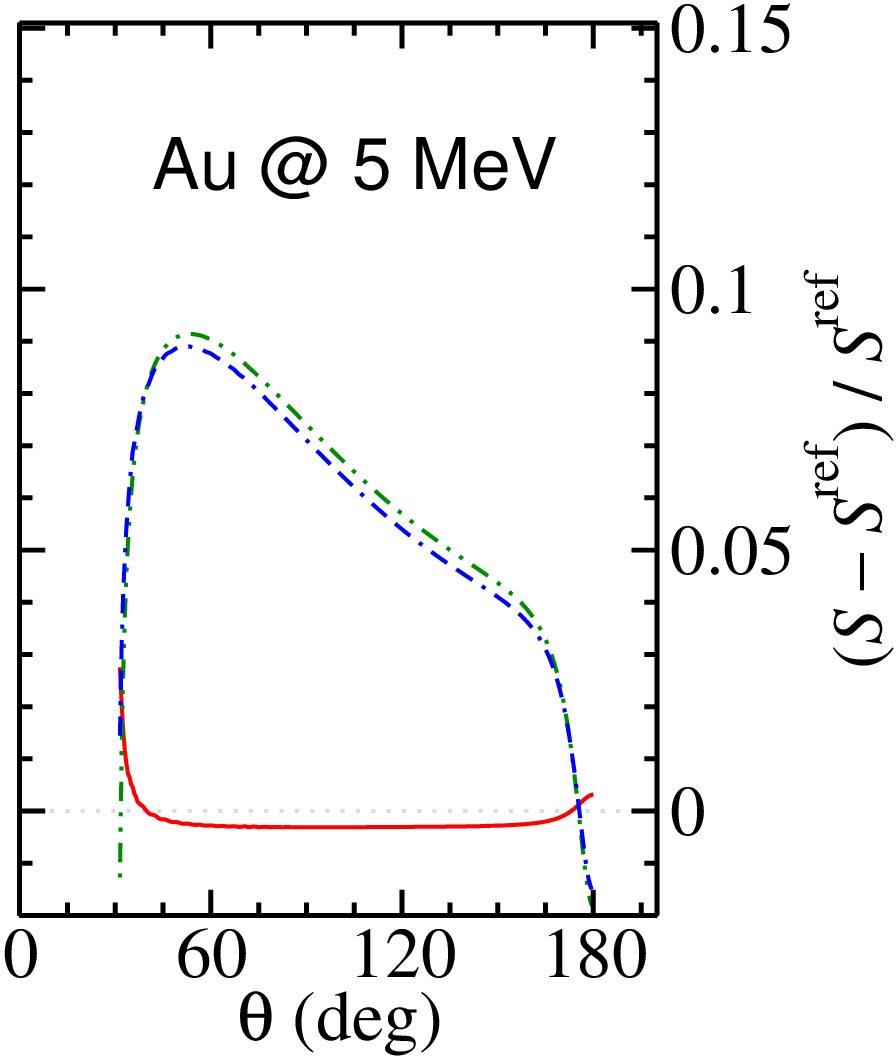}
\caption{Relative change on the Sherman function due to the different approximations with respect to the full calculation are displayed. See text for details.}
\label{sherman-ag}
\end{figure}

In Fig.\ref{sherman-ag} we show the relative change of the Sherman function dependence on the scattering angle for 5 MeV electrons and  ${}^{109}$Ag (left panel) and ${}^{197}$Au (right panel) targets: assuming a point-like nucleus with (double-dot dashed) and without (double-dash dotted) accounting for the presence of atomic electrons; and neglecting atomic electrons and assuming different models describing the finite size of the nuclear charge distribution: Fermi model (full), Helm model (dashed) and SCMF (dash dotted) named G2 \cite{g2}\footnote{Latter two cases only shown for ${}^{109}$Ag since the same situation is found for ${}^{197}$Au.}. Similar results to those obtained for the DCS are also found here. While the effect of accounting or not for the presence of atomic electrons is almost irrelevant for our purposes, the finite size of the nucleus may produce a change on the Sherman function of few \%.   

\section{Conclusions}
\label{conclusions}
Calculations presented here are realistic within a 0.5\% in the Sherman function for the kinematical conditions and targets of interest. The main source of theoretical uncertainties come from radiative corrections which have been shown experimentally to produce about a 0.5\% discrepancy in the Sherman function \cite{steigerwald}. Experiments at different energies and for different nuclei ($Z$) may help understanding the effect of radiative corrections and foster further theoretical studies. 


\begin{acknowledgments}
I would like to acknowledge D. H. Jakubassa-Amundsen for a careful reading of the present manuscript, useful discussions and numerical tests as well as J. M. Grames and C. K. Sinclair for useful discussions.
\end{acknowledgments}

\bibliography{bibliography}

\end{document}